\documentclass[showpacs,aps,print,showkeys,twocolumn]{revtex4}
\usepackage{amsfonts}
\usepackage{amsmath}
\usepackage{amssymb}
\usepackage{graphicx}

\begin{document}

\title{Dicke Phase Transition and Collapse of Superradiant Phase in
Optomechanical Cavity with Arbitrary Number of Atoms }
\author{Xiuqin Zhao}
\affiliation{Institute of Theoretical Physics, Shanxi University, Taiyuan, Shanxi 030006,
China}
\affiliation{Department of Physics, Taiyuan Normal University, Taiyuan, Shanxi 030001,
China}
\author{Ni Liu}
\email{liuni2011520@sxu.edu.cn}
\affiliation{Institute of Theoretical Physics, Shanxi University, Taiyuan, Shanxi 030006,
China}
\author{Xuemin Bai}
\affiliation{Institute of Theoretical Physics, Shanxi University, Taiyuan, Shanxi 030006,
China}
\author{J.-Q. Liang}
\email{ jqliang@sxu.edu.cn}
\affiliation{Institute of Theoretical Physics, Shanxi University, Taiyuan, Shanxi 030006,
China}

\begin{abstract}
We in this paper derive the analytical expressions of ground-state energy,
average photon-number, and the atomic population by means of the
spin-coherent-state variational method for arbitrary number of atoms in an
optomechanical cavity. It is found that the existence of mechanical
oscillator does not affect the phase boundary between the normal and
superradiant phases. However, the superradiant phase collapses by the
resonant damping of the oscillator when the atom-field coupling increases to
a so-called turning point. As a consequence the system undergoes at this
point an additional phase transition from the superradiant phase to a new
normal phase of the atomic population-inversion state. The region of
superradiant phase decreases with the increase of photon-phonon coupling. It
shrinks to zero at a critical value of the coupling and a direct atomic
population transfer appears between two atom-levels. Moreover we find an
unstable nonzero-photon state, which is the counterpart of the superradiant
state. In the absence of oscillator our result reduces exactly to that of
Dicke model. Particularly the ground-state energy for $N=1$ (i.e. the Rabi
model) is in perfect agreement with the numerical diagonalization in a wide
region of coupling constant for both red and blue detuning. The Dicke phase
transition remains for the Rabi model in agreement with the recent
observation.
\end{abstract}

\maketitle

\section{Introduction}

The Dicke model describing $N$ two-level atoms interacting with a
single-mode bosonic field is a paradigm to study fascinating
collective-quantum-phenomena of the light-matter system \cite%
{Dic54,JLJ06,LEB04,DVi04,EBr03,LJo04,HLi73}, the ground state of which
undergoes a quantum phase transition (QPT) by the variation of atom-field
coupling \cite{LEB04,EBr03} in the infinite-$N$ limit. It is believed that
this QPT is possible only when the critical atom-field coupling reaches the
same order of magnitude as the atomic level splitting \cite{Dic54,EBr03}.
The Dicke model has opened an exciting avenue of research in a variety of
context from quantum optics to condensed matter physics since it is a
striking example for the macroscopic many-particle quantum state, which can
be solved rigorously. The experimental realization \cite{BGB10,KMB11} of QPT
from normal phase (NP) to superradiant phase (SP) is a milestone in this
field. This is achieved with a Bose-Einstein condensate (BEC) in an optical
cavity by detecting the photon numbers \cite{BGB10,KMB11}. The QPT of Dicke
model has been well studied \cite{Dic54,EBr03,CLL06} theoretically based on
the variational method with the help of Holstein-Primakoff transformation
\cite{ETB03} to convert the pseudospin operators into a one-mode bosonic
operator in the thermodynamic limit. The ground-state properties were also
revealed in terms of the catastrophe formalism \cite{GNa78}, the coherent
state theory \cite{LZL12,CNL11}, the dynamic method \cite{DEP07,HRi01}, and
the boson expansion \cite{ETB03} as well. Collapse and revival of
oscillations are demonstrated in a parametrically excited BEC in combined
harmonic and optical lattice trap \cite{VermaP}.

The optical cavity coupled with a mechanical oscillator was originally used
to explore the boundaries between classical and quantum mechanics \cite%
{BVT80}. This hybrid system receives renewed interest due to the
experimental progress with the laser cooling of the mechanical mode \cite%
{SDN06,CCI07}, which is a substantial step toward the quantum regime \cite%
{MMT97,MSP03}. The study of such system now becomes a new research branch
known as cavity optomechanics, which is a major resource for implementing
high-precision measurement and quantum-information processing \cite{AKM}.
The micro-engineered mechanical oscillator is coupled with the cavity mode
by the radiation pressure, which generates a nonlinear interaction between
the photon of cavity mode and the phonon of nano-mechanical oscillation. The
influence of mirror motion on the QPT for an optomechanical Dicke model is
investigated by the semi-classical steady-state analysis including also the
dissipation damping \cite{AggarwalN}. In a recent review paper the\textbf{\ }%
cavity quantum electrodynamics is presented for ultracold atoms in optical
and optomechanical\textbf{\ }cavities \cite{DebnathK}. Recently the
variational ground-state and related QPT for a BEC trapped in an
optomechanical cavity were investigated \cite{LLL13} with the
Holstein-Primakoff transformation under the large-$N$ limit. It was also
suggested that this QPT could be observed by measuring the dynamics of
nano-mechanical oscillator \cite{SSF10}.

We in this paper study the ground-state properties of two-level atoms in an
optomechanical cavity by means of the recently developed spin coherent-state
\cite{Rad71,MVe03,AHa12} (SCS) variational method \cite{LZL12,ZLL14}. This
method is not only valid for arbitrary atom number $N$ but also has
advantages to include the inverted pseudospin state ($\Uparrow $), which was
firstly considered \cite{BMS12} in the nonequilibrium dynamics of Dicke
model. The SCS variational method is universal from the $N$-atom Dicke model
to one-atom Rabi model.\textbf{\ }We report the mechanical-oscillator
induced collapse of the SP and the related multiple Dicke phase transitions
for the finite $N$ different from the QPT with $N$ tending to infinity..

\section{Spin coherent-state variational method}

The optomechanical system consists of a high-finesse single-mode optical
cavity of frequency $\omega $ with a fixed mirror and a movable mirror,
which is coupled to a mechanical oscillator \cite{PHH99}. We assume that $N$
two-level $^{87}$Rb atoms with\ transition frequency $\omega _{a}$ are
trapped in the quantized cavity shown schematically in Fig. 1. Although the
QPT has been realized experimentally with an external pump laser \cite%
{BGB10,DEP07,CSD07} we in this paper consider only the simple Dicke\textbf{-}%
model cavity in order to demonstrate the effect of oscillator in a clear
manner. The optomechanical cavity with $N$-atom can be described by the
following Hamiltonian \cite{BGB10,DEP07,ABh09} (with the convention $\hbar $
$=$ $1$):
\begin{equation}
H=H_{DM}+\omega _{b}b^{\dag }b-\frac{\zeta }{\sqrt{N}}\left( b^{\dag
}+b\right) a^{\dag }a\text{,}  \label{1}
\end{equation}%
in which%
\begin{equation}
H_{DM}=\omega a^{\dag }a+\omega _{a}J_{z}+\frac{g}{2\sqrt{N}}\left( a^{\dag
}+a\right) \left( J_{+}+J_{-}\right)
\end{equation}%
is the standard Dicke model Hamiltonian \cite{WHi73,Hio73}. Where $a^{\dag }$
($a$) is the photon creation (annihilation) operator and $b$ ($b^{\dag }$)
is the phonon annihilation (creation) operator of the single vibrational
mode of the nano-oscillator. The ensemble of $N$ atoms is represented by
collective spin operators $J_{z}$, $J_{\pm }$, which satisfy the angular
momentum commutation relations $\left[ J_{+},J_{-}\right] =2J_{z}$ and $%
[J_{z},J_{\pm }]$ $=$ $\pm J_{\pm }$ with eigenvalue $j=N/2$. $g$ denotes
the collective atom-field coupling strength. $\zeta $ is the coupling
constant between the nano-oscillator and cavity mode via radiation pressure
\cite{KV08,MG09,FK09,AGHK10,HWAH11}. A three-body interaction term \cite%
{AKM,SimonG, SantosJP,ChangY} denoted by $\chi (b^{\dag }+b)(a^{\dag
}+a)J_{x}$\ in Hamiltonian (\ref{1}) is removed by specific choice of the
coupling parameters \cite{SantosJP} $\zeta $, $\chi $, since only the
two-body (photon, phonon) radiation-pressure interaction is realized
experimentally \cite{SimonG}.

\begin{figure}[tp]
\includegraphics[width=2.5in]{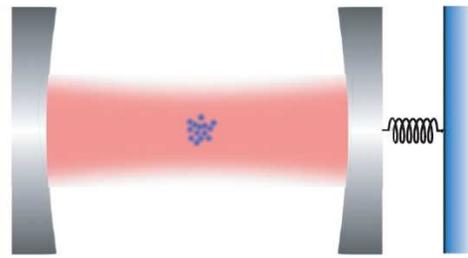}
\caption{Schematic diagram for $N$ atoms in a single-mode optical cavity of
frequency $\protect\omega $. The cavity consists of a fixed mirror and a
movable mirror, which is coupled to a mechanical oscillator.}
\label{fig1}
\end{figure}
We begin with the average of Hamiltonian Eq. (\ref{1}) in boson-operator
states only
\begin{equation}
\overset{\_}{H}\left( \alpha ,\beta \right) =\langle u|H|u\rangle \text{,}
\label{par}
\end{equation}%
where the trial wave-function%
\begin{equation*}
|u\rangle =\left\vert \alpha \right\rangle \left\vert \beta \right\rangle
\text{,}
\end{equation*}%
is considered as the direct product of photon and phonon coherent states $%
a\left\vert \alpha \right\rangle =\alpha \left\vert \alpha \right\rangle $, $%
b\left\vert \beta \right\rangle =\beta \left\vert \beta \right\rangle $. In
this average both photon and phonon operators are replaced by the complex
boson-operator eigenvalues such that $\langle \alpha |a\left\vert \alpha
\right\rangle =\alpha $, $\langle \alpha |a^{\dag }\left\vert \alpha
\right\rangle =\alpha ^{\ast }$; $\langle \beta |b\left\vert \beta
\right\rangle =\beta $, $\langle \beta |b^{\dag }\left\vert \beta
\right\rangle =\beta ^{\ast }$. Moreover we parameterize the complex
eigenvalues $\alpha $, $\beta $ as
\begin{equation*}
\alpha =\gamma e^{i\eta }\text{,\qquad }\beta =\rho e^{i\xi }\text{.}
\end{equation*}%
After the average in the boson coherent-state the Hamiltonian becomes

\begin{equation}
\overset{\_}{H}\left( \alpha ,\beta \right) =(\omega -\frac{2\zeta \rho \cos
\xi }{\sqrt{N}})\gamma ^{2}+\omega _{b}\rho ^{2}+H_{sp}\text{,}
\end{equation}%
in which%
\begin{equation*}
H_{sp}=\omega _{a}J_{z}+\frac{g\gamma \cos \eta \left( J_{+}+J_{-}\right) }{%
\sqrt{N}}
\end{equation*}%
is an effective spin-Hamiltonian. Different from the variational method with
the Holstein-Primakoff transformation, we in the following are going to
diagonalize the spin Hamiltonian $H_{sp}\left( \alpha ,\beta \right) $ in
terms of the SCSs $|\mp \mathbf{n}\rangle $ such that
\begin{equation}
H_{sp}\left( \alpha ,\beta \right) |\mp \mathbf{n}\rangle =E_{sp}^{\mp }|\mp
\mathbf{n}\rangle \text{.}  \label{5}
\end{equation}%
We can visualize $|\mp \mathbf{n}\rangle $ as two eigenstates of a
projection angular momentum operator, $\mathbf{J}\cdot \mathbf{n|\mp n}%
\rangle =\mp j$ $|\mp \mathbf{n}\rangle $, with $\mathbf{n}=\left( \sin
\theta \cos \varphi ,\sin \theta \sin \varphi ,\cos \theta \right) $ being
the unit vector. The directional angles $\theta $ and $\varphi $ as unknown
parameters are to be determined from the eigenstate equation Eq. (\ref{5}).
The SCSs $|\mp \mathbf{n}\rangle $ called respectively the normal ($%
\Downarrow $) and inverted ($\Uparrow $) pseudospin states \cite{BMS12} can
be generated from the extreme Dicke state $\left\vert j,-j\right\rangle $ ( $%
J_{z}$ $\left\vert j,-j\right\rangle $ $=-j\left\vert j,-j\right\rangle $)
with the SCS transformation, $\left\vert \mp \mathbf{n}\right\rangle =U(%
\mathbf{n})\left\vert j,\mp j\right\rangle $. The unitary operator is given
by \cite{ACG72}
\begin{equation}
U(\mathbf{n})=e^{\frac{\theta }{2}(J_{+}e^{-i\varphi }-J_{-}e^{i\varphi })}%
\text{.}  \label{u}
\end{equation}%
In the SCS, spin operators satisfy the minimum uncertainty relation, for
example, $\Delta J_{+}\Delta J_{-}=\left\langle J_{z}\right\rangle /2$ and
therefore it is called the macroscopic quantum state. Applying the unitary
transformation Eq. (\ref{u}) to the eigenvalue equation Eq. (\ref{5}) and
using the unitary transformation relations $\ \ \ \ \ \ \ \ \ \ \ \ \ \ \ \
\ \ \ \ \ \ \ \ \ \ \ $
\begin{equation*}
\left\{
\begin{array}{c}
U^{\dagger }J_{z}U=J_{z}\cos \theta +\frac{1}{2}e^{i\varphi }J_{-}\sin
\theta +\frac{1}{2}e^{-i\varphi }J_{+}\sin \theta \text{,} \\
U^{\dagger }J_{+}U=J_{+}\cos ^{2}\frac{\theta }{2}-e^{i\varphi }J_{z}\sin
\theta -\frac{1}{2}J_{-}e^{2i\varphi }\sin ^{2}\frac{\theta }{2}\text{,} \\
U^{\dagger }J_{-}U=J_{-}\cos ^{2}\frac{\theta }{2}-e^{-i\varphi }J_{z}\sin
\theta -\frac{1}{2}J_{+}e^{-2i\varphi }\sin ^{2}\frac{\theta }{2}\text{,}%
\end{array}%
\right.
\end{equation*}%
it is easy to realize that the SCSs $\left\vert \mp \mathbf{n}\right\rangle $
indeed are the eigenfunctions of effective spin-Hamiltonian $H_{sp}\left(
\alpha ,\beta \right) $ if the following conditions%
\begin{equation}
\left\{
\begin{array}{c}
\frac{\omega _{a}}{2}e^{-i\varphi }\sin \theta +\frac{g\gamma }{\sqrt{N}}%
\cos \eta \left( \cos ^{2}\frac{\theta }{2}-e^{-2i\varphi }\sin ^{2}\frac{%
\theta }{2}\right) =0\text{,} \\
\frac{\omega _{a}}{2}e^{i\varphi }\sin \theta +\frac{g\gamma }{\sqrt{N}}\cos
\eta \left( \cos ^{2}\frac{\theta }{2}-e^{2i\varphi }\sin ^{2}\frac{\theta }{%
2}\right) =0\text{, }%
\end{array}%
\right.  \label{4}
\end{equation}%
are satisfied. The energy eigenvalues are found as
\begin{equation*}
E_{sp}^{\mp }=\mp \frac{N}{2}A\left( \alpha ,\theta ,\varphi \right) \text{,}
\end{equation*}%
in which the parameter function is defined by
\begin{equation*}
A\left( \alpha ,\theta ,\varphi \right) =\omega _{a}\cos \theta -\frac{2g}{%
\sqrt{N}}\gamma \cos \eta \cos \varphi \sin \theta \text{.}
\end{equation*}%
The angle parameters $\theta $ and $\varphi $ can be determined from Eq. (%
\ref{4}). We obtain after a tedious algebra
\begin{equation}
A\left( \gamma \right) =\omega _{a}\sqrt{1+f^{2}(\gamma )}\text{,}
\end{equation}%
which becomes a one parameter function only with%
\begin{equation*}
f(\gamma )=\frac{2g}{\omega _{a}\sqrt{N}}\gamma \text{.}
\end{equation*}%
The total trial wave-function \cite{CNL11,LLM96,CLS04} is a direct product
of the SCSs $\left\vert \mp \mathbf{n}\right\rangle $ and the boson
coherent-states $|u\rangle $
\begin{equation}
\left\vert \psi _{\mp }\right\rangle =|u\rangle \left\vert \mp \mathbf{n}%
\right\rangle \text{.}
\end{equation}%
The variational energy-function in the trial state is evaluated as

\begin{eqnarray}
E_{\mp }\left( \gamma ,\rho ,\xi \right) &=&\langle \psi _{\mp }|H\left\vert
\psi _{\mp }\right\rangle  \notag \\
&=&\omega \gamma ^{2}-\frac{2\zeta \rho \cos \xi }{\sqrt{N}}\gamma
^{2}+\omega _{b}\rho ^{2}  \notag \\
&&\mp \frac{N}{2}A\left( \gamma \right) \text{,}  \label{E}
\end{eqnarray}%
which depends on three variational parameters $\gamma $, $\rho $, and $\xi $%
. The ground state can be determined by the variation of energy function $%
E_{\mp }\left( \gamma ,\rho ,\xi \right) $ with respect to the three
variational parameters. The energy functions Eq. (\ref{E}) are valid for any
atom number $N$ unlike the variational method with the Holstein-Primakoff
transformation, in which large-$N$ limit is required.

\section{Ground state and phase diagram}

The ground state is considered as the variational minimum of energy function
$E\left( \gamma ,\rho ,\xi \right) $ with respect to the variation
parameters $\gamma $, $\rho $, $\xi $. From the usual extremum condition of
the energy function $\partial E/\partial \xi =0$, and $\partial E/\partial
\rho =0$ we find the relation
\begin{equation}
\text{ \ }\rho =\frac{\zeta \gamma ^{2}}{\sqrt{N}}\text{,}  \label{6}
\end{equation}%
and the isolated parameter $\xi $ is determined as $\cos \xi =1$. Replacing
the parameter $\rho $ in the energy function Eq. (\ref{E}) by the relation
Eq. (\ref{6}) the variation-energy becomes a one-parameter function only
\begin{equation}
E_{\mp }(\gamma )=\omega \gamma ^{2}-\frac{\zeta ^{2}}{N\omega _{b}}\gamma
^{4}\mp \frac{N}{2}A\left( \gamma \right) \text{.}  \label{8}
\end{equation}%
The extremum condition of energy function $\partial E_{\mp }/\partial \gamma
=0$ possesses always a zero photon-number solution ($\gamma =0$), which is
called the NP, if it is stable with a positive second-order derivative
\begin{equation}
\frac{\partial ^{2}E_{\mp }(\gamma =0)}{\partial \gamma ^{2}}=2\left( \omega
\mp \frac{g^{2}}{\omega _{a}}\right) >0\text{.}  \label{s}
\end{equation}%
From the stability condition Eq. (\ref{s}) it is easy to find that the NP
state for the normal spin \ ($\Downarrow $) denoted by $N_{-}$ exists only
when
\begin{equation*}
g<g_{c}=\sqrt{\omega \omega _{a}}\text{,}
\end{equation*}%
where $g_{c}$ is the well known critical point \cite{EBr03,ETB03} of the
phase transition between NP and SP in the Dicke model. The zero-photon
solution for the inverted spin ($\Uparrow $) denoted by $N_{+}$ exists,
however, in the whole region of $g$. The mechanical oscillator does not
affect the NP and the critical point $g_{c}$ at all. The extremum condition
of energy function for the nonzero photon solution
\begin{equation*}
\frac{\partial E_{\mp }(\gamma )}{\partial \gamma }=0
\end{equation*}%
becomes effectively a cubic power equation of the variable $\overset{\_}{%
\gamma }^{2}=\gamma ^{2}/N$ seen to be

\begin{equation}
p_{\mp }(\overset{\_}{\gamma })=\omega -\frac{2\zeta ^{2}\overset{\_}{\gamma
}^{2}}{\omega _{b}}\mp \frac{g^{2}}{A(\overset{\_}{\gamma })}=0\text{,}
\label{9}
\end{equation}%
where $A(\overset{\_}{\gamma })=\omega _{a}\sqrt{1+f^{2}(\overset{\_}{\gamma
})}$ with $f(\overset{\_}{\gamma })=2g\overset{\_}{\gamma }/\omega _{a}$.
The Eq.(\ref{9}) can be solved graphically. According to the experimental
parameters \cite{BGB10,KMB11}, we set the atom resonant frequency $\omega
_{a}=$\ $1$\ $MHz$ and the collective atom-field coupling strength $g$
should be in the same order \cite{BGB10,KMB11} to realize the Dicke phase
transition. The frequency of nano-oscillator $\omega _{b}$ ranged from
megahertz to gigahertz \cite{AAU09} and the photon-phonon coupling constant $%
\zeta $\ is of the order of megahertz \cite{AKM}. In the numerical
evaluation the coupling constants, field frequency, and energy are measured
in the unit of atom frequency $\omega _{a}$ throughout the paper. We assume $%
\omega _{b}=10$ in this paper. Fig. 2 displays the plots of polynomial $%
p_{\mp }(\overset{\_}{\gamma })$ and scaled energy
\begin{equation}
\varepsilon _{\mp }(\overset{\_}{\gamma })=\frac{E_{\mp }(\gamma )}{N}
\end{equation}%
as functions of the variable\ $\overset{\_}{\gamma }$ for a given
photon-phonon coupling $\zeta =1$ and various atom-field coupling values to
show the $g$-dependence of the solutions. The second-order derivative of the
energy function
\begin{equation}
\frac{\mathbf{\partial }^{2}\mathbf{\varepsilon }_{\mp }\left( \overset{\_}{%
\gamma }\right) }{\mathbf{\partial }\overset{\_}{\gamma }^{2}}=2\left(
\omega \mathbf{-}\frac{6\zeta ^{2}\overset{\_}{\gamma }^{2}}{\omega _{b}}%
\mathbf{\mp }\frac{g^{2}\omega _{a}^{\frac{1}{2}}}{A^{\frac{3}{2}}(\overset{%
\_}{\gamma })}\right)  \label{st}
\end{equation}%
serves as the stability condition of the nonzero photon state. The solution
of energy extremum condition Eq. (\ref{9}) denoted by $\gamma _{s}^{-}$ is a
stable state, which is the root of the polynomial $p_{-}(\overset{\_}{\gamma
})$ with a positive slope of the curve [Fig. 2(a2)] and, therefore, the
positive second-order derivative of the energy function $\partial ^{2}%
\mathbf{\varepsilon }_{-}\left( \overset{\_}{\gamma }\right) /\partial
\overset{\_}{\gamma }^{2}>0$. The state $\gamma _{s}^{-}$, which is a local
minimum of the energy function [Fig. 2(b2)], is called the SP in the phase
diagram. The higher value solution denoted by $\gamma _{us}^{-}$ [Fig.
2(a2)] with a negative slope (namely the negative second-order derivative of
the energy function) is a unstable state corresponding to the local maximum
of the energy function (b2). For the fixed photon-phonon coupling $\zeta =1$%
, two solutions move close to each other with the increase of the atom-field
coupling $g$ and finally coincide at a critical value $g_{t}=1.763$ (a3)
called the turning point, where the energy function becomes a flexing point
(b3). Above this point the SP no longer exists. The unstable state $\gamma
_{us}^{-}$ extends also to the NP region below the critical point $g_{c}$
seen from Fig. 2(a1), (b1). The variation of the SP-state $\gamma _{s}^{-}$
with the photon-phonon coupling $\zeta $ is shown in the\textbf{\ }Fig. 3%
\textbf{\ }for a fixed atom-field coupling $g=1.5$. The two solutions $%
\gamma _{s}^{-}$ and $\gamma _{us}^{-}$ also move close to each other with
the increase of $\zeta $. Particularly when the photon-phonon coupling
reaches a critical value $\zeta =1.203$ the two states $\gamma _{s}^{-}$, $%
\gamma _{us}^{-}$ coincide and the energy curve becomes a flexing point
corresponding to the turning point $g_{t}=1.5$ seen from Fig. 4.

\bigskip
\begin{figure}[tp]
\includegraphics[width=3.0in]{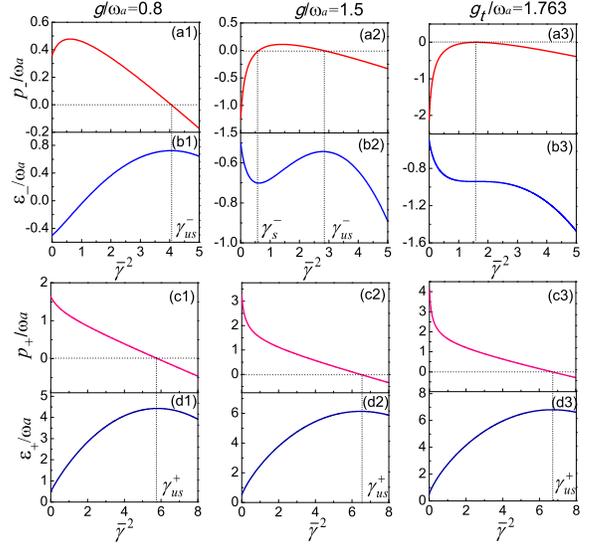}
\caption{(Color online) Graphical solution of the energy extremum equations $%
p_{-}(\protect\overset{-}{\protect\gamma })=0$ (upper panel), $p_{+}(%
\protect\overset{-}{\protect\gamma })=0$ (lower panel) for given
photon-phonon coupling\ $\protect\zeta =1.0$ and different atom-field
couplings $g=0.8$ (a1)(c1), $1.5$ (a2)(c2), $1.763$ (a3)(c3). The
corresponding average-energy curves are plotted in (b1)(d1), (b2)(d2),
(b3)(d3). $\protect\gamma _{s}^{-}$ denotes stable nonzero photon solution
of normal spin ($\Downarrow $) with a local energy minimum and $\protect%
\gamma _{us}^{-}$ is the unstable one (energy maximum). There is no any
stable solution for the inverted spin ($\Uparrow $) but the unstable one
denoted by $\protect\gamma _{us}^{+}$.}
\label{fig.2}
\end{figure}

\begin{figure}[tp]
\includegraphics[width=2.5in]{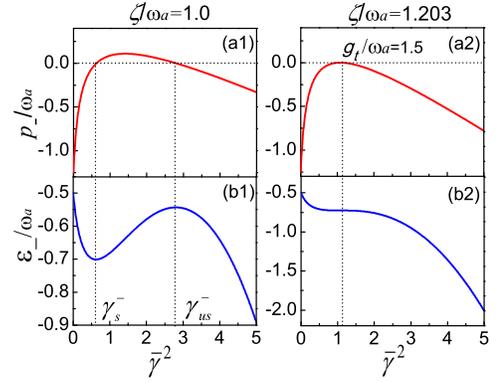}
\caption{(Color online) $\protect\zeta $-dependence of the nonzero photon
solution of the energy extremum equation $p_{-}(\protect\overset{\_}{\protect%
\gamma })=0$ (a) and the corresponding energy curve $\protect\varepsilon %
_{-} $ (b) with $\protect\zeta =1.0$ (1), $1.203$ (2), for the fixed
atom-field coupling $g=1.5$. With the increase of $\protect\zeta $ the
stable and unstable solutions move close to each other and finally coincide
at the turning point seen from (a1) (a2). }
\label{fig.3}
\end{figure}
For the inverted spin ($\Uparrow $) we only find the unstable nonzero photon
state denoted by $\gamma _{us}^{+}$ in the system considered [Fig. 2(c)].
Fig. 4 is the phase diagram of $g$-$\zeta $ plane obtained from the extremum
Eq. (\ref{9}) and stability condition Eq. (\ref{st}) with $\omega =1$. The
mechanical oscillator does not affect the NP and the critical point $g_{c}$
between NP and SP. The SP is, however, bounded by the turning-point line $%
g_{t}$, beyond which the SP collapses by the resonant damping of the
nano-oscillator. In the region above the turning-point line $g_{t}$ the zero
photon solution $N_{+}$ of Eq. (\ref{9}) still exists for the inverted spin (%
$\Uparrow $) and becomes the ground state. This NP with the state $N_{+}$ is
denoted by the phase notation $NP(N_{+})$ in the phase diagram (Fig. 4).
Thus the system undergoes an additional phase transition at the turning
point $g_{t}$ from the SP to the $NP(N_{+})$.

\begin{figure}[tp]
\includegraphics[width=2.5in]{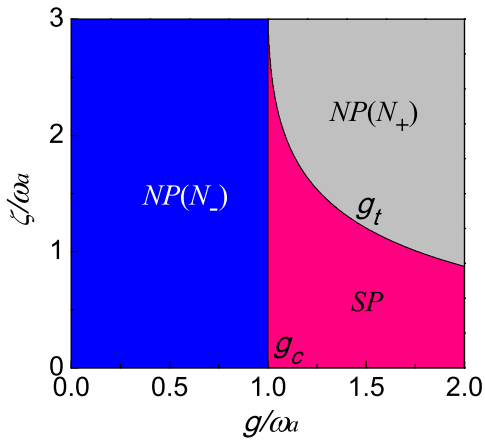}
\caption{Phase diagram in $g$-$\protect\zeta $ plane. The SP is restricted
by two boundary lines of $g_{c}$ and $g_{t}$. When $\protect\zeta =0$ the
phase transition reduces exactly to that of Dicke model from the $NP(N_{-})$
to the SP at the critical point $g_{c}$. The region of SP extends from $%
g_{c} $ to infinity. On the other hand the region of SP decreases with the
increase of $\protect\zeta $ and disappears completely when $\protect\zeta %
=3.0$ resulting in the direct transfer between two NP-states $N_{-}$ and $%
N_{+}$.}
\label{fig.4}
\end{figure}

The turning point shifts back to the lower value direction of atom-field
coupling $g$ with the increase of photon-phonon coupling $\zeta $ and the SP
disappears completely at the value $\zeta =3$.

\section{ Collapse of the superradiant phase and atomic population transfer}

The nano-oscillator induces collapse of the SP due to resonant damping. To
see detail of the effect we study the variations of average photon number,
energy and atomic population with respect to the coupling constants $g$, $%
\zeta $. The mean photon number in the SP is obviously
\begin{equation}
n_{p}=\frac{\left\langle \alpha |a^{\dag }a|\alpha \right\rangle }{N}=\left(
\gamma _{s}^{-}\right) ^{2}\text{,}
\end{equation}%
here $\gamma _{s}^{-}$ means the corresponding $\overset{\_}{\gamma }$-value
in the SP of normal spin ($\Downarrow $) as indicated in Figs. 2 and 3. The
atomic population difference is evaluated from average of the collective
pseudospin operator in the normal spin state $|-\mathbf{n\rangle }$
\begin{eqnarray}
\Delta n_{a} &=&\frac{\left\langle -\mathbf{n}|J_{z}|-\mathbf{n}%
\right\rangle }{N}=\frac{\left\langle j,-j|U^{\dag }(\mathbf{n})|J_{z}U(%
\mathbf{n})\mathbf{|}j,-j\right\rangle }{N}  \notag \\
&=&-\frac{1}{2\sqrt{1+f^{2}(\gamma _{s}^{-})}}\text{,}
\end{eqnarray}%
which becomes the known value in the NP, that $\Delta n_{a}(\gamma
_{-}=0)=-1/2$. While the atomic population difference in the state $N_{+}$
is $\Delta n_{a}(\gamma _{+}=0)=1/2$ indicating the atomic population
inversion. The average of phonon number is proportional to the square of
photon number that

\begin{equation}
n_{b}=\frac{\left\langle \beta |b^{\dag }b|\beta \right\rangle }{N}=\frac{%
\zeta ^{2}}{\omega _{b}^{2}}n_{p}^{2}\text{.}
\end{equation}

\begin{figure*}[t]
\includegraphics[width=7.0in]{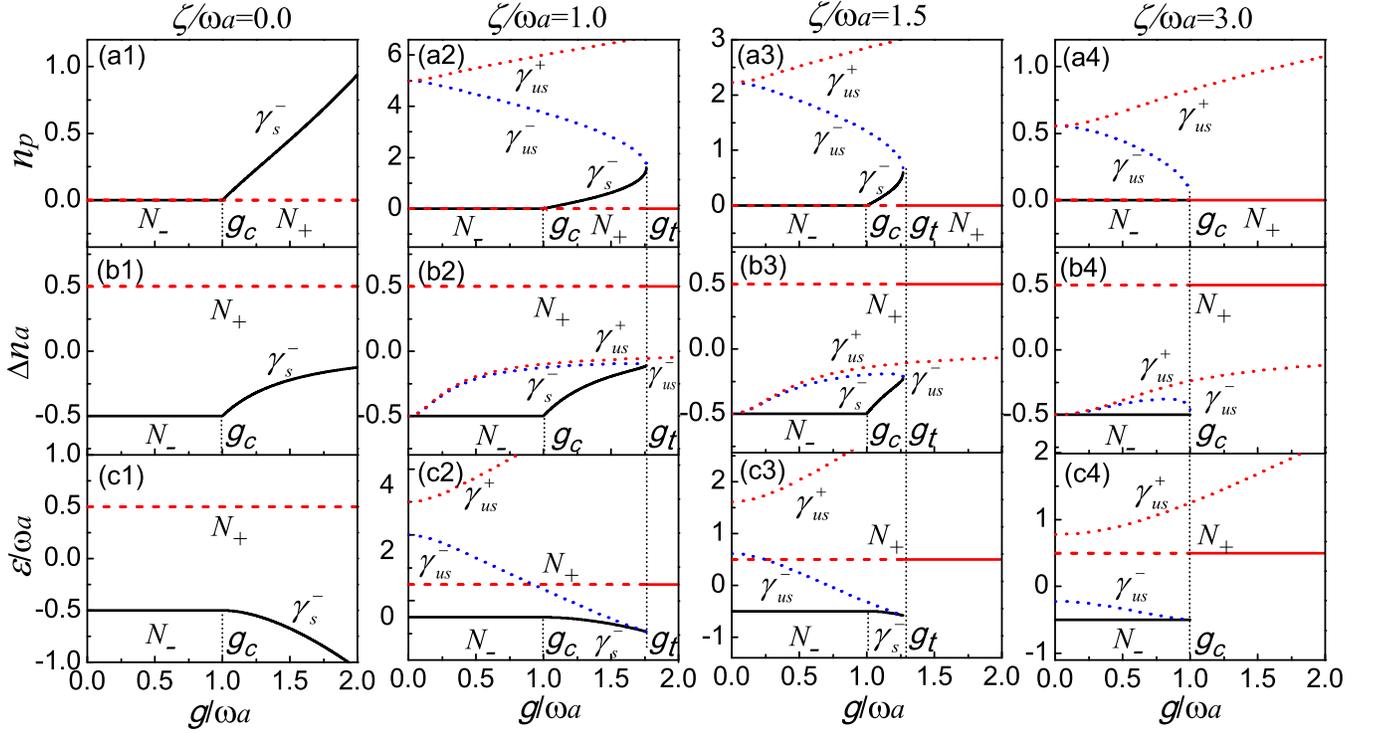}
\caption{(Color online) The photon-phonon coupling dependence of average
photon number $n_{p}$ (a), atomic population difference $n_{a}$ (b) and
average energy $\protect\varepsilon $ (c). The typical phase transition of
Dicke model ($\protect\zeta =0$) is shown in (1), where $g_{c}$ is the
critical point between $NP$($N_{-}$) (black solid line) and $SP$($\protect%
\gamma _{s}^{-}$) (black solid line). The SP collapses at the turning point $%
g_{t}$ (2), (3) and the zero photon state $N_{+}$ becomes the ground state
(red solid line). The system undergoes the second phase transition from SP
to the $NP(N_{+})$. When $\mathbf{\protect\zeta =}3.0$ (4) the SP disappears
and we see the direct atomic population transfer between two states $N_{-}$
and $N_{+}$ [or spin flip between normal ($\Downarrow $) and inverted ($%
\Uparrow $) spins]. Beyond the turning point $g_{t}$ the nonzero photon
state turns back to an upper branch $\protect\gamma _{us}^{-}$ (blue dotted
line), which is unstable. $\protect\gamma _{us}^{+}$ (red dotted line)
denotes unstable state of inverted spin.}
\label{fig.5}
\end{figure*}

The average photon number can be considered as an order parameter with $%
n_{p}>0$ denoting the SP and $n_{p}=0$ for the NP. We plot in Fig. 5 the $%
n_{p}$ curves for different phonon-photon couplings $\zeta $ compared with
the normal Dicke model ($\zeta =0$) shown in Fig. 5(1) to demonstrate the
nano-oscillator induced effect. Below the critical point $g_{c}$ we have the
bistable zero photon states $N_{\mp }$ , in which the $N_{-}$ (black solid
line) with lower energy [Fig. 5(c)] is the ground state and $N_{+}$ is the
excited state (red dash line). The phase transition from $NP(N_{-})$ to SP
of the nonzero photon state $\gamma _{s}^{-}$ takes place at the critical
point $g_{c}$. In this region the zero photon state $N_{+}$ (red dash lines)
is still excited state with higher energy than superradiant state $\gamma
_{s}^{-}$ seen from Fig. 5(c1-c3). The SP collapses at the turning point $%
g_{t}$ and the zero photon state $N_{+}$ of inverted spin ($\Uparrow $) (the
atomic population inversion state) shown in Fig. 5(b) becomes the ground
state (red solid line). An additional phase transition from SP to the $%
NP(N_{+})$ appears at the turning point $g_{t}$. This multiple phase
transitions generated by the mechanical oscillator does not exist at all in
the usual Dicke model [Fig. 5(1)]. With the increase of photon-phonon
coupling $\zeta $ the region of SP is suppressed and the turning point $%
g_{t} $ shifts back to the lower value direction of $g$ [Fig. 5(2,3)]. The
SP disappears completely at a critical value $\zeta =3$ and the phase
transition becomes the transition from the $NP(N_{-})$ to $NP(N_{+})$ within
the same phase of zero order-parameter $\gamma =0$ [Fig. 5(a4)] but
different average energy [Fig. 5(c4)] and atomic population [Fig. 5(b4)]. We
observe a interesting phenomenon of atomic population transfer (or spin
flip) between two atom levels (b4). Since the atomic population inversion
plays a important role in the laser physics, the controllable population
transfer certainly has technical applications. Beyond the turning point $%
g_{t}$ the nonzero photon state turns back to a upper branch state $\gamma
_{us}^{-}$ (blue dotted lines) so that we have a dual state for one value of
$g$. This is similar to the optical bistability, however, the state $\gamma
_{us}^{-}$ is unstable in the considered system. The unstable state $\gamma
_{us}^{+}$ (red dotted lines) of inverted spin ($\Uparrow $) possesses more
higher values of photon-number, atomic population, and average energy as
well.

It may be worthwhile to emphasize again that the SCS variational method \cite%
{LLL13} has advantage to avoid the thermodynamic limit and the Dicke phase
transition demonstrated in this paper is valid for arbitrary atom number $N$%
. We in the following section compare our results with those in the
literature.

\section{Dicke phase transition for Rabi model}

The Rabi model describes a two-level atom in a single-mode cavity \cite%
{Rab36} and has been widely used in atomic, optic, and condensed matter
physics \cite{SZu97}. The validity of the model has been experimentally
verified in various systems \cite{RBH01,LBM03,NPT01,FLM10,Lon11,CLO12,LWe11}%
. In the absence of the mechanical oscillator ($\zeta =0$) the Hamiltonian
Eq. (\ref{1}) of optomechanical cavity reduces exactly to that of the Rabi
model for $N=1$. The energy function has a simple form \cite%
{LZL12,LLM96,CLS04} (we in this section consider only the normal spin ($%
\Downarrow $) and neglect the subscript "-" for the simplicity). The energy
function becomes simply
\begin{equation}
E\left( \gamma \right) =\omega \gamma ^{2}-\frac{A(\gamma )}{2}\text{.}
\end{equation}%
The extremum equation $\partial E(\gamma )/\partial \gamma =0$ gives rise to
the stable zero-photon solution $\gamma =0$, namely the NP, below the
critical point%
\begin{equation*}
g_{c}=\sqrt{\omega \omega _{a}}.
\end{equation*}%
The photon number in the SP above $g_{c}$ is found as
\begin{equation}
n_{p}=\langle \alpha |a^{\dag }a|\alpha \rangle =\gamma _{s}^{2}=\frac{1}{4}%
\left( \frac{g^{2}}{\omega ^{2}}-\frac{\omega _{a}^{2}}{g^{2}}\right) ,
\end{equation}%
which is exactly the same as the average photon number $n_{p}$ =$\langle
\alpha |a^{\dag }a|\alpha \rangle /N$ in the normal Dicke model \cite%
{ETB03,CLL06,LZL12} with $N$ atoms. The ground state energy
\begin{equation}
E\left( g\right) =\left\{
\begin{array}{c}
-\frac{\omega _{a}}{2}\text{, }g\leq g_{c}\text{,} \\
-\frac{\omega }{4}\left( \frac{g^{2}}{\omega ^{2}}+\frac{\omega _{a}^{2}}{%
g^{2}}\right) \text{, }g>g_{c}\text{.}%
\end{array}%
\right.  \label{e}
\end{equation}%
equals also to the average energy of Dicke model \cite{ETB03,CLL06,LZL12}.
We present the energy curves obtained from the analytic formula of Eq. (\ref%
{e}) and the numerical diagonalization together in Fig. 6. We see the
perfect coincidence in a wide range of coupling value and for both the red
and blue detuning. The small deviations come out in the large detuning as
shown by the dashed lines. The deviation decreases with the increase of
photon-number $n$ of Fock state in the numerical diagonalization. The phase
transition indeed exists in the Rabi model in agreement with the recent
observation \cite{HPP15}.
\begin{figure}[t]
\includegraphics[width=3.0in]{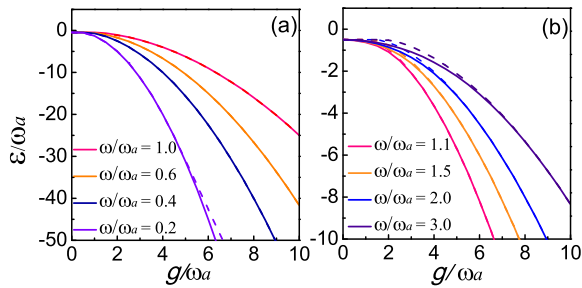}
\caption{(Color online) The energy curves obtained from both analytic
formula Eq.(\protect\ref{e}) and the numerical diagonalization with Fock
state up to $300$-photon for the red detuning (a) and blue detuning (b). The
deviation of numerical-diagonalization line from the analytic result is
indicated by the dashed line.}
\label{fig.6}
\end{figure}

\section{Conclusion and discussion}

The SCS variational method is a universal tool in the study of ground-states
for various atom-cavity systems with arbitrary atom number $N$. The Dicke
phase transition is a common phenomenon independent of the atom number $N$.
The only assumption is that the cavity field is in the coherent state or in
other words the macroscopic quantum state. The nano-oscillator coupled with
the cavity mode by radiation-pressure does not affect the phase transition
boundary from the NP to SP. However the SP state collapses at a turning
point due to the resonant damping of the oscillator. As a consequence of
collapse the zero-photon state-$N_{+}$ of inverted spin ($\Uparrow $)
becomes a ground-state and an additional phase transition from the SP to $%
NP(N_{+})$ takes place at the turning point. Particularly a direct atomic
population-transfer between two atom-levels is realized by the manipulation
of photon-phonon coupling. This novel observation may have technical
applications in laser physics. In the absence of mechanical oscillator ($%
\zeta =0$) our result recover the phase transition of Dicke model. The phase
transition remains for Rabi model with $N=1$ in agreement with the recent
observation \cite{HPP15}. This conclusion is also verified by the
quantitative agreement of average energy with the result of numerical
diagonalization in a wide region of coupling constant for both the red and
blue detuning. The dual nonzero-photon states similar to the optical
bistability are also found, however, the upper branch of state is unstable
in the system considered.

\section{Acknowledgments}

This work was supported by the National Natural Science Foundation of China
(Grant Nos. 11275118, 11404198, 91430109), and the Scientific and
Technological Innovation Programs of Higher Education Institutions in Shanxi
Province (STIP) (Grant No. 2014102), and the Launch of the Scientific
Research of Shanxi University (Grant No. 011151801004), and the National
Fundamental Fund of Personnel Training (Grant No. J1103210). The natural
science foundation of Shanxi Province (Grant No. 2015011008).

\bigskip

\end{document}